\documentclass[pdflatex,sn-mathphys-num]{sn-jnl}
\usepackage{makecell}

\usepackage{graphicx}%
\usepackage{multirow}%
\usepackage{amsmath,amssymb,amsfonts}%
\usepackage{amsthm}%
\usepackage{mathrsfs}%
\usepackage[title]{appendix}%
\usepackage{xcolor}%
\usepackage{textcomp}%
\usepackage{manyfoot}%
\usepackage{booktabs}%
\usepackage{algorithm}%
\usepackage{algorithmicx}%
\usepackage{algpseudocode}%
\usepackage{listings}%
\usepackage{hyperref}
\usepackage{siunitx}

\theoremstyle{thmstyleone}%
\theoremstyle{thmstyletwo}%

\theoremstyle{thmstylethree}%

\newcommand{\revision}[1]{{\color{black}{{#1}}}}

\begin{document}

\title[Training framework for Kidney Ureteroscopy Exploration Assessment]{Automated Assessment of Kidney Ureteroscopy Exploration for Training}


\author*[1]{\fnm{Fangjie} \sur{Li}}\email{fangjie.li@vanderbilt.edu}
\author[2]{\fnm{Nicholas} \sur{Kavoussi}}
\author[2]{\fnm{Charan} \sur{Mohan}}
\author[1]{\fnm{Matthieu} \sur{Chabanas}}
\author[1]{\fnm{Jie Ying} \sur{Wu}}

\affil*[1]{\orgdiv{Department of Computer Science}, \orgname{Vanderbilt University}, \orgaddress{\street{2301 Vanderbilt Place}, \city{Nashville}, \postcode{37235}, \state{TN}, \country{USA}}}

\affil[2]{\orgdiv{Department of Urology}, \orgname{Vanderbilt University Medical Center}, \orgaddress{\street{1211 Medical Center Drive}, \city{Nashville}, \postcode{37232}, \state{TN}, \country{USA}}}

\abstract{
\textbf{Purpose:} Kidney ureteroscopic navigation is challenging with a steep learning curve. However, current clinical training has major deficiencies, as it requires one-on-one feedback from experts and occurs in the operating room (OR). Therefore, there is a need for a phantom training system with automated feedback to greatly \revision{expand} training opportunities.

\textbf{Methods:} We propose a novel, purely ureteroscope video-based scope localization framework that automatically identifies calyces missed by the trainee in a phantom kidney exploration. 
\revision{We use a slow, thorough, prior exploration video of the kidney to generate a reference reconstruction. Then, this reference reconstruction can be used to localize any exploration video of the same phantom.}

\textbf{Results:} In 15 exploration videos, a total of \revision{69 out of 74 calyces} were correctly classified. We achieve $< 4\ mm$ camera pose localization error. Given the reference reconstruction, the system takes 10 minutes to generate the results for a typical exploration (1-2 minute long).

\textbf{Conclusion:} We demonstrate a novel camera localization framework that can provide accurate and automatic feedback for kidney phantom explorations. We show its ability as a valid tool that enables out-of-OR training without requiring supervision from an expert.
}


\maketitle

\section{Introduction}

In ureteroscopic kidney stone removal operations, up to $20\%$ of the patients require a second operation due to missed stones~\cite{brain2021natural}. This is partly due to the challenging nature of navigating through the kidney collecting framework~\cite{yamany2015ureterorenoscopy}, which requires precise endoscopic manipulation and knowledge of the kidney's anatomy to ensure that every kidney cavity, called a calyx, is fully visited. 

Accurately navigating the kidney has a steep learning curve. 
Unfortunately, training opportunities are limited because the current training paradigm relies on one-on-one, apprenticeship-style guidance during operating room (OR) cases, which are subject to significant time and safety constraints~\cite{arora2010impact}. 
Additionally, trainees often only receive limited verbal feedback at the end of a case~\cite{bai2020user}, which is based primarily on an expert's subjective judgment. In this work, we aim to improve ureteropscopy training by introducing an automated objective feedback mechanism when exploring kidney phantoms.

In prior work~\cite{AcarAyberk2025NNAR}, we introduced anatomically accurate phantoms that can be used for training outside of the OR. However, training on these phantoms still requires one-on-one guidance from an expert. Although electromagnetic tracking used in~\cite{AcarAyberk2025NNAR} can provide automated feedback on exploration completeness, they also increase the hardware cost and complexity. Automatic assessment of exploration completeness without additional hardware may facilitate the adoption of ureteroscopy training on phantoms. 

Computer vision-based methods such as Simultaneous Localization and Mapping (SLAM) and Structure from Motion (SfM) show great promise in reconstructing the anatomical scene and localizing scope poses through video input alone~\cite{oliva2023orb,schmidt2024tracking}. However, their performance is susceptible to the quality of exploration videos~\cite{widya20193d}. This leads to frequent failures when using the algorithms on poor quality exploration videos with high motion blur, such as those generated by trainees learning to use the endoscope.

In this paper, we propose a novel, purely ureteroscope video-based framework that measures the trainee's exploration coverage of a kidney, providing automatic feedback. It identifies calyces missed by the trainee. The framework does not require additional hardware other than a consumer-grade computer. We overcome the challenges with existing methods for ureteroscope-based reconstruction by using a \revision{slow and thorough reference} exploration of the kidney phantom to \revision{build its reference 3D reconstruction}. This reconstruction acts as a \revision{reusable} prior, simplifying the pose localization problem for the \revision{challenging normal-speed query} exploration videos \revision{by trainees}. The reconstruction can be re-used for any exploration videos of the same phantom. The code base will be publicly available upon the acceptance of the paper.




\revision{\section{Related Work} }



\textbf{Simulation Models for Uretroscopic Training:} 
Traditionally, surgical training followed an apprenticeship model, but simulation-based training is gaining traction as they add training opportunities, provides objective performance metrics, and have demonstrated educational effectiveness~\cite{brunckhorst2015simulation}. Physical bench-top models exist for urology applications, such as the Uro-Scopic Trainer (Limbs \& Things Ltd., Bristol, UK), Scope Trainer (Mediskills Ltd., Edinburgh, UK), and the adult ureteroscopy trainer (Ideal Anatomic Modelling, MI). They mimic the anatomical structure and texture with high fidelity. Their realism are confirmed through clinical user studies, and users' score in the system correlated with the user's experience~\cite{matsumoto2001novel,brehmer2002validation,white2010validation}. Despite their realism, none of these physical models have automatic feedback on task performance.

\revision{\textbf{3D Reconstruction in Surgical Applications:}
3D reconstruction or camera position tracking in medical research field is progressing quickly due to its ability to provide intraoprative information of the surgical scene to improve guidance precision~\cite{schmidt2024tracking, zhang20213d}.} In colonoscopy, such reconstruction algorithms have been developed for ensuring full exploration coverage~\cite{zhang20213d}, similar to our goal in ureteroscopy. 3D reconstruction with ureteroscopy is relatively underexplored. Maza, et al.~\cite{oliva2023orb} adapted a SLAM algorithm for ureteroscopy applications, \revision{adding image preprocessing and adapting a new image features detection algorithm. Nonetheless, the system can still lose track under rapid motions. Acar et al.~\cite{acar2024towards} performed SfM reconstruction of patient and the CT rendered ureteroscopy using different methods, of which hloc~\cite{sarlin2019coarse} produced most robust results. Overall, endoscope images with poor image quality such as motion blur present significant challenge for these system to function robustly~\cite{widya20193d}. }


\section{Methodology}

\subsection{Framework Description}
\label{sec:Framework_Description}
\begin{figure}
\includegraphics[width=0.99\textwidth]{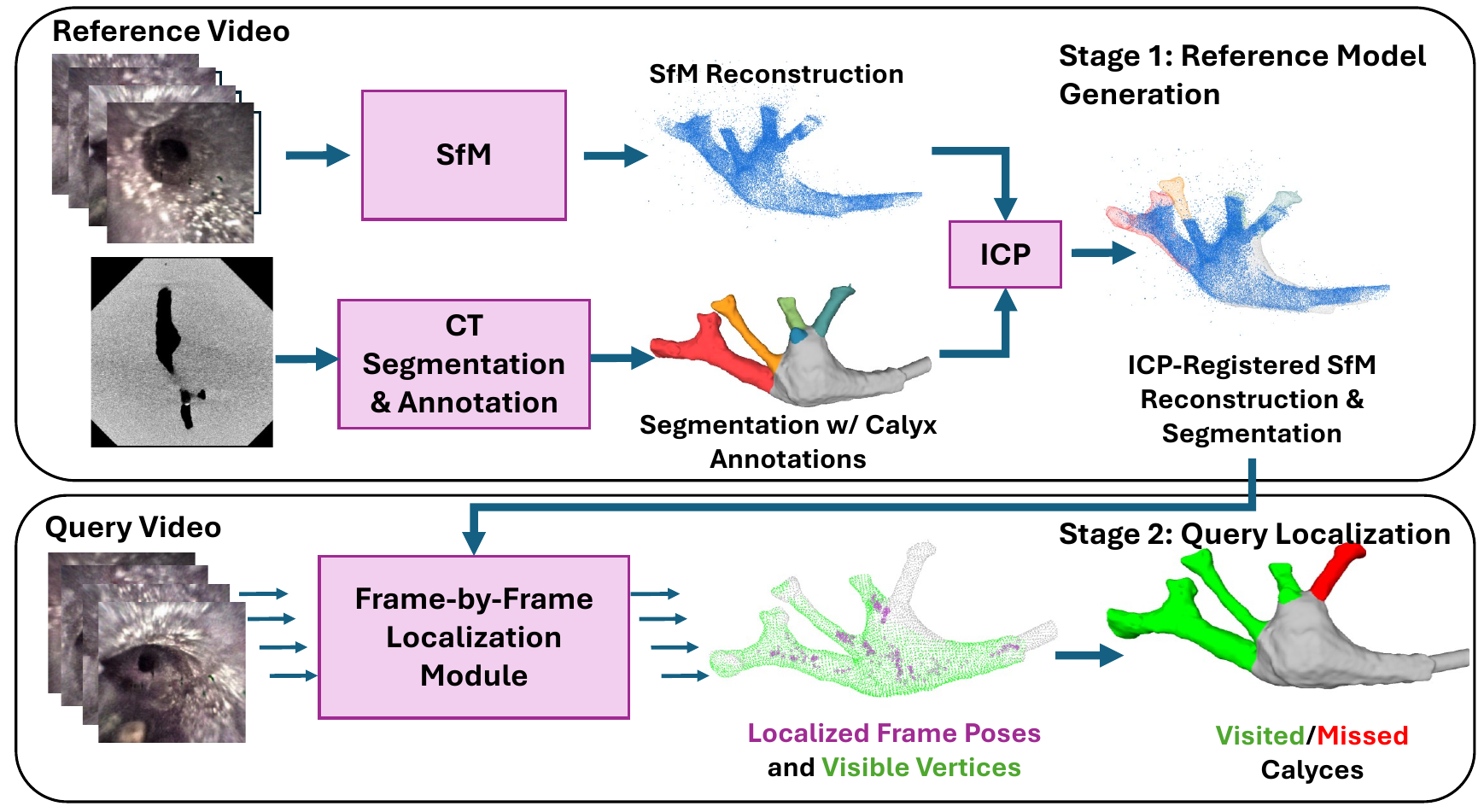}
\caption{The overall workflow of the framework. In stage one, a reference \revision{reconstruction} is generated and registered to the CT segmentation. This \revision{reconstruction} is then reused for all query exploration videos, which are localized in stage two. For each video, calyces are marked as visited/not visited.} \label{fig:system_figure}
\end{figure}

Our framework consist of two stages \revision{(Fig.~\ref{fig:system_figure})} . In the first stage, a reference \revision{3D} reconstruction of the phantom's collecting framework is generated with an SfM algorithm and \revision{two slow and thorough} reference exploration videos~\cite{acar2024towards, acar2025monocular}. \revision{The reference reconstruction consists of a 3D point cloud of the kidney collecting system, and endoscope poses of the slow reference exploration frames localized within the point cloud.} The reconstruction is then manually registered to the computed tomography (CT) segmentation of the phantom. In the second stage, the framework receives a \revision{normal-speed} query video of a phantom exploration \revision{from each trainee}. The framework localizes the query video frames based on the reference reconstruction from stage one. This allows the framework to classify kidney calyces as visited or missed. The framework then displays localization information to the user via an annotated CT segmentation. For any phantom, the \revision{reference reconstruction (and hence, the slow exploration)} from stage one can be reused for localizing any number of exploration videos in stage two. Additionally, this two-stage approach allows for the calyx-level localization of videos that cannot be reconstructed on their own \revision{due to their limited quality}.


\textbf{Stage 1: Reference Model Generation:}
\revision{For each phantom, we use two slow, thorough explorations as reference videos} of the kidney phantom’s collecting system. \revision{Because we use an SfM pipeline invariant to image order, we simply concatenated the frames from two videos. This results} in a total of 1500–2300 frames per phantom\revision{, from which we generate a reference reconstruction of the collecting system}. \revision{To ensure maximum coverage and high image quality, we ask experts to perform the explorations.} \revision{We employ an SfM pipeline based on the hloc toolkit, because in our prior endoscopic reconstruction work~\cite{acar2024towards,acar2025monocular}, hloc met our robustness needs and compared favorably to alternative SfM pipelines on similar data. In principle, any SfM/SLAM approach that is reliable for the slow videos could be sufficient for our framework.} The pipeline uses NetVLAD \cite{arandjelovic2016netvlad} for covisible image retrieval, ALIKED \cite{zhao2023aliked} for feature detection, LightGlue for feature matching, and COLMAP \cite{schoenberger2016sfm} for multi-view 3D reconstruction. The phantom is scanned with computer tomography (CT) and manually segmented \revision{by the authors}. The reference reconstruction is registered to the CT segmentation using Iterative Closest Point (ICP) with manual initialization. The individual calyces in the CT segmentation are also manually annotated, to enable per-calyx visitation classification. The renal pelvis and ureter are not annotated as exploration in those regions is trivial for kidney exploration. \revision{The slow videos are only required once at this stage for each phantom to generate a reference reconstruction, and all query videos of the same phantom can reuse the same reference reconstruction.}

\textbf{{Stage 2a: Query Video Localization}}
Using the reference reconstruction and CT segmentation, we can localize challenging, \revision{normal-speed} query exploration videos of the same phantom. For each query frame, a two-stage image retrieval process is performed. First, NetVLAD retrieves candidate covisible reference images. Second, to improve localization robustness, we use a local-feature based match filter to remove falsely matched reference images. Feature extraction and matching are performed with ALIKED and LightGlue. Outlier matches are filtered through RANSAC-based essential matrix estimation, and reference frames with inlier matches, inlier ratio below thresholds are rejected. The remaining high-confidence reference frames matches are then used to determine the camera pose associated with the query frame. Afterwards, we deploy further spatio-temporal consistency filtering to remove any incorrectly localized frames. First, all query frames that are localized outside of the segmentation mesh are rejected. Second, query frames are filtered based on its distance to the last localized frame. We set a dynamic distance threshold based on the difference in time from the last localized frame (ex. frames further separated in time may be further separated in distance). 



\textbf{Stage 2b: Calyx Visit Score Computation}
For each localized query frame, we obtain its 6 degrees-of-freedom (DoF) poses localized against the CT segmentation. Combined with the intrinsics parameters of the ureteroscope, we render the view from the estimated camera pose within the CT segmentation mesh (Fig.~\ref{fig:qualitative_localization_results}). We then mark the vertices of the CT segmentation visible from the localized pose, through ray casting. Given the entire query video, we aggregate all the vertices that were viewed.

We then compute the visitation score for each calyx individually. For each calyx
, we determine the total number of vertices belonging to that calyx and the subset of those vertices that were visited. The visitation score is then defined as the ratio of \revision{viewed} vertices to the total number of vertices in the calyx. \revision{We then set a} threshold, $VS_{thd}$, where calyces with a score higher than $VS_{\text{thd}}$ are considered thoroughly visited. Overall, this produces a \revision{binary} classification output for all segmented calyces in the phantom, where each calyx is marked as visited or missed. The main renal pelvis and entrance are not marked, as they represent trivial exploration targets. 

\revision{
\textbf{Parameters and Filter Threshold Settings}
All parameters in the framework are set globally (i.e., not per model or per video), based on heuristics developed in prior studies and cross-validation, and fixed prior to evaluation.

In Stage 2a, RANSAC filter parameters for rejecting invalid image and feature matches were set heuristically based on our prior experience with feature-based endoscopic reconstruction \cite{acar2024towards, acar2025monocular} and a small subset of frames from three videos. Such RANSAC-based filtering is standard practice in SfM pipelines. These parameters were then fixed and applied uniformly to all other videos to avoid overfitting.

Also in Stage 2a, a distance-based filter is applied to filter out incorrectly localized poses. We first selected one of the three videos mentioned above, and estimated the scope velocity based on the reconstruction. We then set the filter threshold heuristically and conservatively at approximately 135 mm/s—well above normal exploration speeds for our phantoms, which are no longer than 150 mm in length.

For the visitation threshold $VS_{thd}$ in stage 2b, which controls the final output accuracy, we performed 5-fold cross-validation on the 15 trainee videos, with each fold containing three randomly sampled videos. For each fold $f$, visitation scores were computed for all calyces in the train-set videos. Based on manual expert annotations, calyces were divided into visited and unvisited sets, yielding two score distributions: $\{S^f_{visited}\}$ and $\{S^f_{non\ visited}\}$. To best separate the two classes, the visitation threshold for fold $f$ is then defined as the midpoint between the mean scores of the two sets:
\begin{equation}
    VS^f_{thd} = \frac{mean\left(\{S^f_{visited}\}\right) + mean\left(\{S^f_{non\ visited}\}\right)}{2}
\end{equation}

This choice provides a simple, unbiased decision boundary between the two score distributions. 
To improve robustness, the 5-fold cross-validation was repeated five times with different random seeds. 
}


\subsection{Experiments}

\textbf{Video Data Collection:}
\label{sec:video_data_processing}
We used anatomically accurate silicone phantom based on patient CT \revision{of kidneys with normal anatomy. The phantoms CT are easy to segment as they were homogeneous in material}. These phantoms were fabricated following the method described in \cite{AcarAyberk2025NNAR}\revision{, with BegoStone fake kidney stones (Bego, USA) inserted}. In total, four phantoms were included and we collected data in two separate experimental sessions.

\revision{In a first experiment, two expert surgeons performed, for each phantom, two thorough explorations at slow speed and one exploration at normal speed. The slow explorations provide high quality, comprehensive scope of the kidney and are used for generating reference reconstructions (stage 1 of the method). The normal-speed explorations serve as validation query videos to quantitatively evaluate stage 2, i.e., that frames collected at normal speed can be localized with respect to the reference reconstruction.}
For each phantom, one slow and one normal-speed exploration were electromagnetically (EM) tracked to provide ground truth camera poses for evaluation. An EM sensor was fixed at the tip of the ureteroscope and tracked using the Aurora EM tracking system (North Digital Inc, Canada).

\revision{In a second experiment, we collected exploration videos conducted at normal-speed by four surgical trainees, over a multi-month period. Each trainee explored the four phantoms, under the guidance of an expert surgeon. For one trainee, we failed to record one exploration, leaving us with 15 trainee query videos. The videos are used as query and localized in stage 2 of the method. EM tracking was not employed, because attaching the EM sensor noticeably increases the ureteroscope diameter (scope diameter: 2 mm, sensor diameter: 1 mm), and interferes with normal scope handling, reducing the fidelity of the simulated ureteroscope experience for trainees.} The trainees, with two to four years of postgraduate experience, were guided by an expert either verbally (standard clinical practice) or using a previously developed AR-based eye-gaze guidance system~\cite{atoum2025sight}. \revision{All trainees are residents from the urology department of the Vanderbilt University Medical Center, USA.}



\revision{Videos were recorded at 30 FPS} and lasted between 1 to 2 minutes. \revision{A stride of 2 was applied.}
\revision{Scope camera intrinsics were calibrated using a ChArUco board.} Data was recorded with approval from the hospital’s Institutional Review Board (IRB 231997).

\revision{\textbf{System Setup} We used a PC with an independent Graphics Card (RTX 4090) for all the computations.}

\textbf{Reference Reconstruction Accuracy:}
We \revision{measure if} the SfM reconstruction point cloud is accurate relative to the CT segmentation, so that camera poses localized against this reconstruction are accurate against the real anatomy shape. After registering the reference point cloud to the CT segmentation, we compute \revision{the mean Euclidean distance from the point cloud to the segmentation mesh (single-sided chamfer distance). We also compute 99 percentile Hausdorff distance, which highlights outliers in the reconstruction.}


Additionally, we compute the reconstruction coverage and the reprojection error. The reconstruction coverage is defined as the percentage of CT point cloud points that have a corresponding point in the reference reconstruction within $1\ mm$ distance. The reprojection error \revision{is the pixel-space distance between an observed 2D feature point and the projected location of its corresponding 3D point under the estimated camera pose and intrinsics}.

\revision{Lastly, we evaluate the accuracy of SfM pose localizations of the EM tracked reference video frames.} We use 10\% of the localized reference frame poses \revision{as fiducials} to estimate the transformation between the reconstruction and EM tracker coordinate frame \revision{origins robustly}. With this transform, we then register the remaining reference frame poses \revision{as targets}. We compute the mean Euclidean distance between the transformed SfM localizations and EM tracked \revision{ground truth}.

\revision{\textbf{Camera Pose Localization Accuracy:}}
\revision{For EM-tracked normal-speed exploration videos from the first experiment, localized using stage 2 of the framework, we quantitatively evaluate pose localization accuracy. The same EM-to-reconstruction transformation described above is applied, and the mean Euclidean distance to the EM-tracked ground-truth poses is computed.}

\revision{EM tracking is not available for trainee explorations. We then perform} qualitative, manual assessment of camera localization accuracy. \revision{For each frame,} we generate a rendered view of the CT segmentation using the camera intrinsics and \revision{estimated} pose (Fig.~\ref{fig:qualitative_localization_results}). Then, we manually review the CT renders side-by-side with the actual query frames, to confirm that that they represented the same anatomical positions.





\textbf{Visitation Classification Accuracy:} A challenge with SfM-based localization is that it can fail to localize frames, due to the lack of salient visual features. \revision{Therefore, we assess if the framework can capture the overall exploration path, even with missed frames.} Two independent reviewers annotated explored and missed calyces for each query video. Where annotations differed, they discussed and came to agreement. These binary annotations are compared against the framework’s output.

\section{Results}
\begin{figure}[tb]
\includegraphics[width=0.99\textwidth]{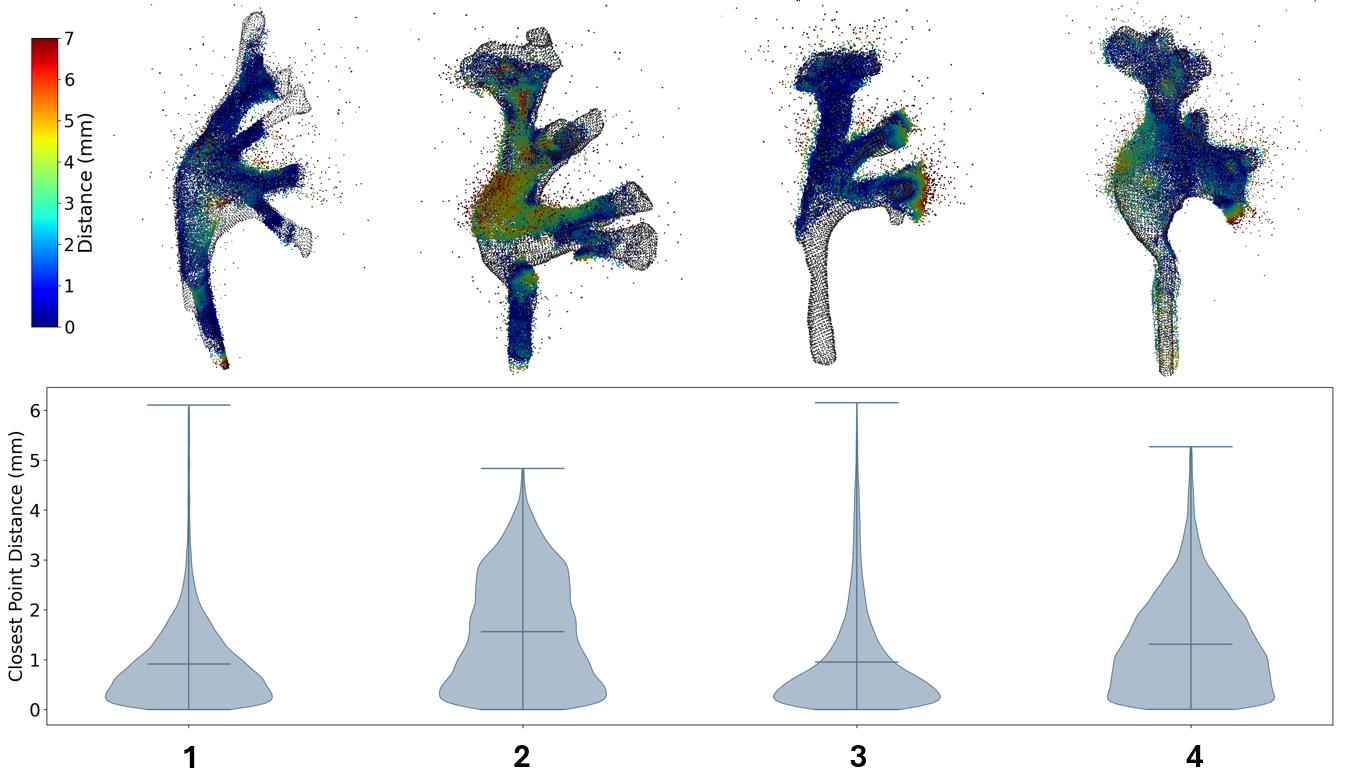}
\caption{The accuracy of the SfM reconstruction point cloud compared to the CT volume. The violin plot is plotted with the 99 percentile data, as the outliers, though numerically extreme, has almost no impact on the CT registration and the subsequent pose localization.} \label{fig:reconstruction_result}
\end{figure}

\textbf{Reference Reconstruction Accuracy:}
As shown in Table~\ref{tab1} \revision{and Fig.~\ref{fig:reconstruction_result}}, all phantom reconstructions had mean \revision{Euclidean} point distances of $<2\ mm$, with a standard deviation of also under $2\ mm$, suggesting good reconstruction accuracy. \revision{All reconstruction had 99\% percentile Hausdorff distance below $6.5\ mm$, suggesting small outliers. Notably}, the reconstruction does not cover the entirety of the CT segmentation. Nonetheless, for phantom 1-3, all calyces had a significant portion of its tubular structure reconstructed \revision{(Fig.~\ref{fig:reconstruction_result})}. \revision{For phantom 4, the lower most calyces are partially reconstructed, due to their hard-to-reach geometry.}

\begin{table}
\centering
\caption{Reference Model Point Cloud CT Registration Error} \label{tab1}
\begin{tabular}{|c|c|c|c|c|}
\hline
Phantom ID &  \revision{\makecell{Mean Euclidean\\Distance (mm)}} & \revision{\makecell{99\% Hausdorff\\Distance (mm)}} & Coverage (\%) & Projection Error (px)\\
\hline
1 &  1.0 $\pm$ 1.8 & \revision{6.1} & 58.19 & 1.19\\
\hline
2 & 1.7 $\pm$ 1.8& \revision{4.8} & 43.17 & 1.14\\
\hline
3 & 1.1 $\pm$ 1.9& \revision{6.2} & 63.76 & 1.32\\
\hline
4 & 1.4 $\pm$ 1.6& \revision{5.3} & 66.30 &1.26\\
\hline
\end{tabular}
\end{table}

\textbf{Quantitative Reconstruction and Localization Camera Pose Accuracy:}
In Table~\ref{tab2}, we display the camera position error of EM tracked reference and query videos. All errors are below $4\ mm$. 



\begin{table}
\centering
\caption{The Camera Pose Localization Accuracy of Expert Recordings}\label{tab2}
\begin{tabular}{|c|c|c|c|}
\hline
Phantom ID & $Video_{\text{ref}\_em}$ Position Error (mm) & $Video_{\text{query}\_em}$ Position Error (mm) \\
\hline
1 & 2.7 $\pm$ 2.0 & 2.6 $\pm$ 1.6 \\
\hline
2 & 2.9$\pm$ 1.7 & 3.5 $\pm$ 1.9 \\
\hline
3 & 3.3 $\pm$ 1.7 & 3.8 $\pm$ 1.7\\
\hline
4 &  3.0 $\pm$1.7 & 3.2 $\pm$ 2.00 \\
\hline
\end{tabular}
\end{table}

\begin{figure}[tb]
\includegraphics[width=0.99\textwidth]{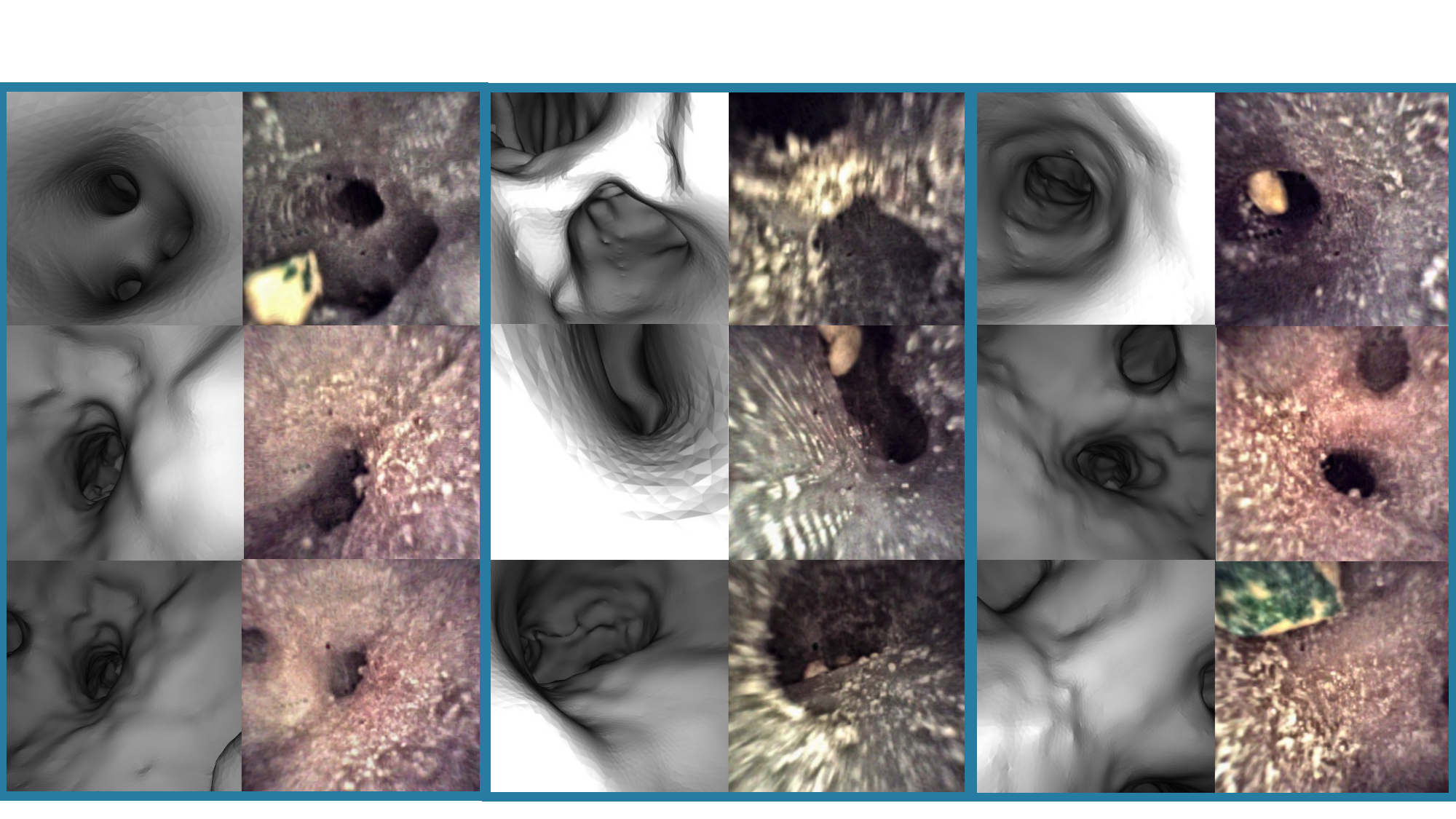}
\caption{A random selection of rendered vs real ureteroscope image Pairs. (Note: the kidney stones are not in the CT view)} \label{fig:qualitative_localization_results}
\end{figure}

\textbf{Qualitative Localization Accuracy:}
In Fig.~\ref{fig:qualitative_localization_results}, we display a random selection of the rendered vs real ureteroscope image pairs, from query trainee videos and their \revision{estimated} localization. In all frames, the corresponding anatomical landmarks are clearly identifiable. 

\textbf{Visitation Classification Accuracy:}
\label{sec:visitation_classification_results}
\revision{In the cross-validation study, an average of 69} out of 74 calyces were correctly classified. \revision{The corresponding classification accuracy is $92.8 \% \ (\text{CI: }91.6\%-94.0\%)$. The mean visitation thresholds are $VS_{thd}^f = 0.45 \pm 0.06$}.

\begin{figure}[tb]
\centering
\includegraphics[width=0.8\textwidth]{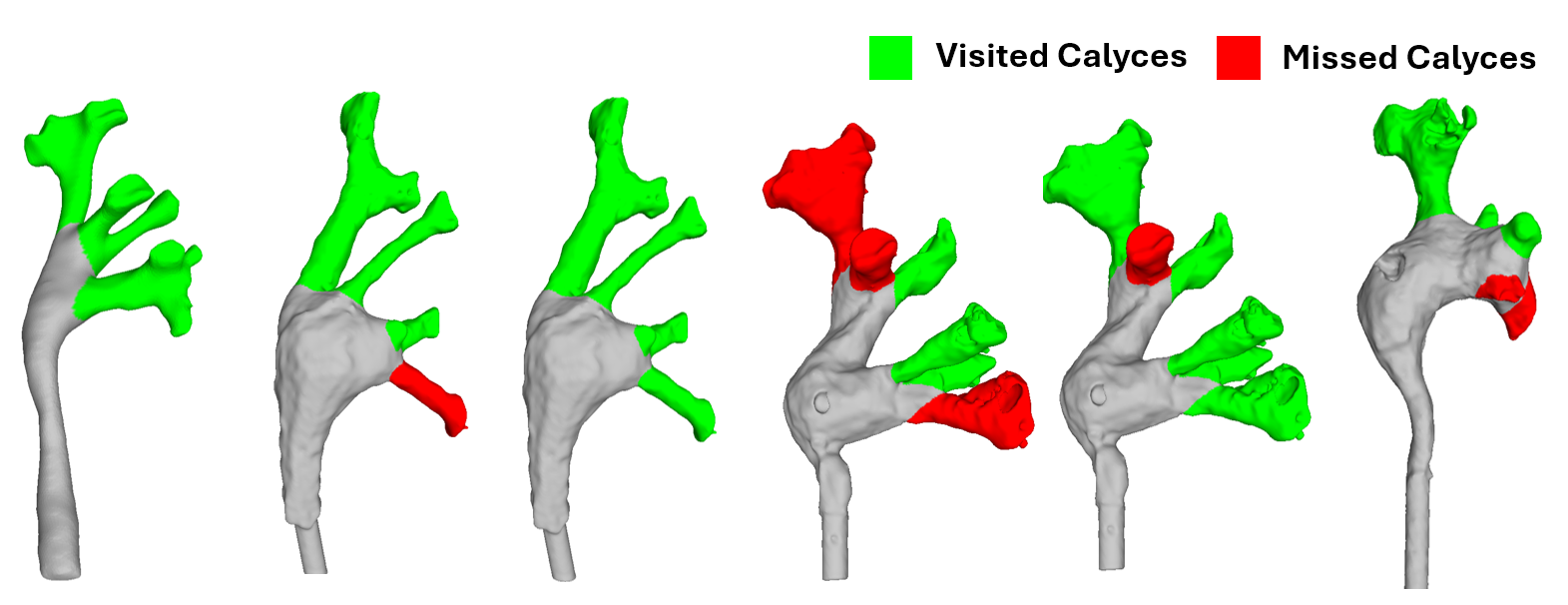}
\caption{Five example cases where the framework accurately identifies visited/missed calyces} \label{fig:visitation_successful}
\end{figure}

\begin{figure}[tb]
\includegraphics[width=0.9\textwidth]{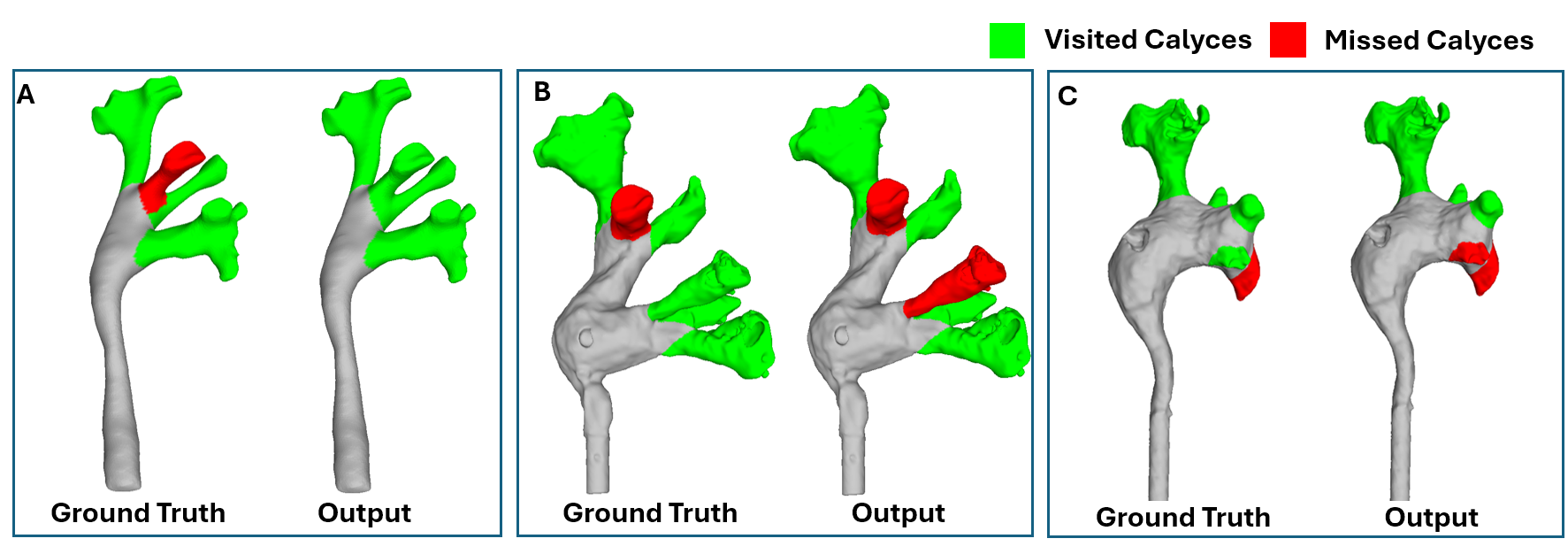}
\caption{\revision{Example cases where} the framework misidentified one calyx in each case. A: Phantom 2. B: Phantom 3. C: Phantom 4.} \label{fig:visitation_failed}
\end{figure}

In Fig.~\ref{fig:visitation_successful}, we display examples of the correct visitation outputs. In Fig.~\ref{fig:visitation_failed}, we display three failed cases, where the framework misclassified one calyx in each. 


\revision{\textbf{System Runtime:}
Generating the reference model took around 40 minutes for each phantom, given 1500-2000 reference frames (after striding). At stage 2, A query video with 1000 frames (after striding) took around 10 minutes. This promises a semi-real-time evaluation tool for phantom training. }

\section{Discussion}
\revision{\subsection{Stage One - Reference Reconstruction Quality}}
The reference models are geometrically accurate with small outliers, given the small \revision{Euclidean distance and Hausdorff distance}.

The major anatomical structures, especially the calyces, correspond well to the CT segmentation. 



\revision{
The clinically relevant regions, particularly the calyces, are well reconstructed. The relatively low numerical reconstruction coverage is mainly due to missing entry-point and posterior renal pelvis regions, which are generally not clinically relevant. Some calyces lack deep/distal portions due to limited parallax from predominantly forward motion, but most of each calyx is reconstructed, supporting query pose localization.
In phantom 4, two lower calyces are incompletely reconstructed because their geometry is hard to reach, requiring near $\SI{180}\degree$ ureteroscope bending. This has two implications for localization: first, such regions are often inaccessible to trainees as well, limiting their practical impact on visitation classification; second, they represent edge cases where framework-based classification (and, to a lesser extent, binary human annotations) become ambiguous.}


\revision{\textbf{Reconstruction Camera Pose Accuracy:}} The mean reference pose errors are under $4\ mm$ (Table~\ref{tab1}) for all phantoms. The error is adequately low for \revision{the geometry}, as the calyces have a diameter of around $10\ mm$, and depths of over $20\ mm$.


\revision{
\subsection{Stage Two - Pose Localization Accuracy}}
\revision{Quantitatively}, the mean query pose errors are all under $4\ mm$ (Table~\ref{tab1}), which is again adequately low for visitation classification, \revision{given the anatomy dimensions mentioned above}. 

\revision{Qualitatively,} the rendered and real image pairs consistently depict the same anatomical structures in correct relative positions, demonstrating the framework’s precision. There are only minor \revision{position} or orientation inaccuracies in \revision{few} cases. This is acceptable for categorical classification of calyx visitation, which do not require highly accurate geometric localization. 



\revision{\subsection{Visitation Classification Accuracy:}}
\revision{Across cross-validation trials, the framework’s visitation classifications closely match expert annotations. In 15 trainee videos, the average accuracy is 92.8\%, with 69 of 74 calyces correctly classified, indicating the framework provides robust automated training feedback. The visitation thresholds are relatively stable ($VS_{thd}^f = 0.45 \pm 0.06$}) across folds, suggesting robustness.


However, \revision{the failure cases listed in figure~\ref{fig:visitation_failed} highlight some limitations} of the current systems. In Fig.~\ref{fig:visitation_failed}-A, wrong classification happens because the user only briefly glances at a calyx\revision{, which the human annotator marks as missed}. As the current \revision{framework} analyzes each frame independently without considering view duration, this leads to a "visited" classification. An estimation of \revision{view duration and scope trajectory} could resolve this limitation. 


\revision{In Fig.~\ref{fig:visitation_failed}-B, a visited calyx is marked as missed, due to the highly erratic motion the user used to reach the calyx, wchich renders almost all frames there blurred.  The current framework is generally robust for normal-speed videos, because it can localize individual high-quality frames that cannot be reconstructed on their own, even if most frames are blurred and of low quality. However, if the entire video sequence exploring a calyx is of low quality, this remains a challenge.}
A retrieval method based on higher-level geometric or lumen shape features \cite{tian2024bronchotrack} may be explored \revision{for their improved robustness over feature-based methods}. 

\revision{Lastly, some cases in phantom 4 (Fig.~\ref{fig:visitation_failed}-C) had visited calyces marked as missed. As discussed, these are hard-to-reach edge cases with incomplete reconstructions, reflected by the fact that no trainee fully explored the lower calyces.}

\subsection{Use Cases of Visitation Classification Results}

\revision{The primary use case of the proposed framework is to provide automatic, unsupervised feedback to trainees, for example through visual summaries of calyces visitation (Fig.~\ref{fig:visitation_successful}). Future human-computer-interface focused studies are needed to evaluate the effect of alternative feedback modalities. Beyond direct feedback, visitation classification could serve as a quantitative metric for skill assessment, provided that a meaningful correlation with trainee skill level or guidance type can be established. The focus of this work was to show that the proposed framework can localize challenging query frames robustly classify calyx visitation. With that in mind, and given the limited number of trainees in the dataset, no correlation was observed between visitation classification outcomes and trainee skill level, or the guidance type. Future studies with a larger cohort are needed to evaluate these potential relationships more rigorously.}






\revision{The system also can reveal cohort-level trends by identifying anatomies missed across users.} For example, all user missed a small calyx in phantom three, and two out of four users missed the lower-most calyx in phantom \revision{four}. This can highlight \revision{difficult-to-access anatomies, informing trainee guidance.}

\revision{Finally, the framework can have potential utility as a preoperative planning tool.} As the phantom can be made from patient CTs, surgeons \revision{may} explore the phantom \revision{prior to} a surgery. The {visitation analysis} can then help identify calyces that are easy to miss\revision{. This may allow surgeons to better plan }for {challenging anatomies and }potentially improve surgical outcomes.


\section{Conclusion}
In conclusion, we introduce a novel approach for identifying missed calyces with high accuracy from exploration videos of kidney phantoms. The approach allows for automated training feedback on kidney exploration. This can reduce the amount of manual supervision in surgical training, and opens up a viable venue for out-of-OR training. Additionally, it can also be a useful tool for surgical planning, providing surgeons with better understanding of easy-to-miss calyces prior to the real surgery, potentially leading to better surgical outcomes.

\section*{Acknowledgments}
This study was partially supported by the NIBIB of the NIH Grant 1R21EB035783.


\bibliography{references}

\end{document}